\documentclass[prl,twocolumn,floatfix,superscriptaddress]{revtex4}
\usepackage{dcolumn,amsmath}
\usepackage[dvips]{graphicx}
\usepackage{bm}
\usepackage{hyperref}

\newcommand{\cm}{\ensuremath{{\rm cm}}$^{-1}$}

\begin{document}
\title{Relativistic coupled cluster calculations 
of spectroscopic and chemical properties for element 120}

\author{L.V.\ Skripnikov}\email{leonidos239@gmail.com}
\affiliation{Federal state budgetary institution ``Petersburg Nuclear Physics  Institute'', Gatchina, Leningrad district 188300, Russia}
\affiliation{Institute of Physics, Saint Petersburg State University, Saint Petersburg, Petrodvoretz 198904, Russia}
\author{N.S.\ Mosyagin}
\affiliation{Federal state budgetary institution ``Petersburg Nuclear Physics  Institute'', Gatchina, Leningrad district 188300, Russia}
\date{2 May 2012}
\author{A.V.\ Titov} 
\affiliation{Federal state budgetary institution ``Petersburg Nuclear Physics  Institute'', Gatchina, Leningrad district 188300, Russia}
\date{6 May 2012}

\begin{abstract}
The coupled cluster calculations with accounting for relativistic effects to
study spectroscopic and chemical properties of element 120 (E120) are performed.
Similar calculations for Ba are also done and they are in a good agreement with
the experimental data. Dissociation energies of diatomic X-H and X-Au molecules,
where X=E120, Ba, are calculated; for E120 they are found to be $1.5\div2$
times smaller than those for Ba.
\end{abstract}

\maketitle

%========================================================================
\section{Introduction}

At present all the relatively long-lived superheavy elements (SHEs)
up to the seventh period of the Periodic table
   with exception of element 118
were synthesized in FLNR~JINR (see
\cite{Oganessian:09a, Oganessian:10a} and references) and their synthesis was
mainly confirmed in LBNL \cite{Stavsetra:09} and GSI \cite{Gates:11a}.  The new
challenge for nuclear physics is the synthesis of the elements of the eight
period.

According to its position in the Periodic table, element 120 (E120) is supposed
to be the $s$ element and homologue of Ba and Ra. The earlier obtained ground
state configuration of E120 at the Dirac-Fock level, $7s^27p^68s^2$, is
confirmed in recent calculation \cite{Dinh:08} using the correlation potential
method.  Relativistic stabilization of the valence $s$ orbital
\cite{Schwerdtfeger:98aa, Kaldor:04ba} leads in general to higher excitation
energies from this shell and, as a consequence, to weakening the bond strengths in
its chemical compounds.  Therefore, E120 can resemble the noble gas rather than
the alkaline earth element.  Provided that E120 is synthesized, its
properties can be investigated experimentally by the gas-phase chromatography
method. Such scheme was successfully employed in FLNR~JINR \cite{Eichler:07a} and
GSI \cite{Gates:11a} for E112 and E114 where gold served as adsorbent.
Theoretical calculations are required for planning similar experiments with
E120 on the gold surface.

We have performed calculations of excitation energies of the E120 atom and its
cation, E120$^+$, as well as of spectroscopic properties of the E120H and E120Au
molecules compared to the analogous systems with Ba, for which some experimental
data are available.

%---------------------------
\section{Computational method}

To study properties of E120, the generalized relativistic effective core
potential (GRECP) method \cite{Mosyagin:10a} was employed.  For relativistic
correlation treatment, the fully-relativistic Fock-space coupled cluster code
with single and double cluster amplitudes (FS-RCCSD) \cite{MolRCCSD,Kaldor:04ba}
was applied.  To evaluate corrections on enlargement of the basis set and
higher cluster amplitudes, the \textsc{cfour} \cite{CFOUR} and \textsc{mrcc}
\cite{Kallay:1,Kallay:3} codes were used.

%--------------------------------------
\subsection{GRECP generation and atomic calculations}

The GRECP for E120 was generated in the framework of the present work.  To check
performance of the GRECP method for such a heavy element, relativistic
correlation calculations on the E120 atom and its cations were carried out. We
used four-component calculations with Dirac-Coulomb-Breit (DCB) Hamiltonian and
Fermi nuclear charge distribution ($A=304$) as the reference study.  This
Hamiltonian takes into account the great bulk of the relativistic effects
including the relativistic corrections to the Coulomb interaction between
electrons. The $1{-}5s$, $2{-}5p$, $3{-}5d$, and $4{-}5f$ shells were frozen
from the E120$^{2+}$ ground state calculation. The 28 electrons (occupying the
$6spd,7spd,8sp$ shells) were correlated in the [12,14,14,13,12,11] basis set of
$6{-}17s_{1/2}$, $6{-}19p_{1/2}$, $6{-}19p_{3/2}$, $\ldots$ $6{-}16h_{11/2}$
numerical spinors localized in the same radial space region as the
$6spd,7spd,8sp$ spinors.
%  due to appropriately chosen (highly-)charged cationic states of the atom used
%  for generating the virtual orbitals.
The correlations were taken into account with the help of the FS-RCCSD method.
The closed-shell ground state of E120$^{2+}$ was the reference state and the
Fock-space scheme was
\begin{equation} {\rm E120}^{2+} \rightarrow {\rm E120}^+ \rightarrow {\rm
E120}, \end{equation}
with electrons added to the $8s_{1/2}$, $8p_{1/2}$, $8p_{3/2}$, $7d_{3/2}$, and
$7d_{5/2}$ spinors (the relativistic configurations corresponding to the
nonrelativitic $8p^2$, $7d^2$, and $8p^1 7d^1$ ones were excluded from the model
space). 

The leading configurations and terms for the lowest-lying states of the E120
atom and its cations are presented in the first and second columns of
table~\ref{All-el}. The results of the semiempirically-fitted CI/MBPT
calculations with Dirac-Coulomb Hamiltonian~\cite{Dinh:08,Dinh:08a} are compared
with our FS-RCCSD results in the third and fourth columns. They are in a
reasonable agreement.
%m: variant The main differences are probably due to neglecting the triple
% cluster amplitudes in the FS-RCCSD method.
%t: variant The main differences are probably due to neglecting the triple
% cluster amplitudes for the valence electrons within the FS-RCCSD method and, in
% smaller extent, by the MBPT2 level of treatment of the $7p$ shell in
% \cite{Dinh:08}.
The absolute errors due to decrease in the number of the correlated electrons,
the neglect of a finite nuclear size or Breit interactions are listed in the
last four columns.  One can see that a finite nuclear size should be taken into
account, whereas the correlations with the $6spd$ electrons and Breit
interactions can be neglected for the accuracy within
2~kcal/mol\,$\approx$\,700~\cm\ for one-electron excitations.

\begin{table}[!h] \caption{Transition energies (TE) from DCB/FS-RCCSD
calculations of the lowest-lying states of the E120 atom and its cations for 28
correlated  electrons with Fermi nuclear model and Breit interactions taken into
account. Absolute errors of reproducing the TE in the different versions of the
four-component calculations. All values are in cm$^{-1}$.  } \label{All-el}
\begin{tabular}{llrrrrrr} 
\hline 
\hline 
\multicolumn{2}{l}{Number of corr.el.:} &  & 28
& 20 & 10   & 28    & 28   \\ \hline &         & TE    & TE
&\multicolumn{4}{c}{Abs.errors}\\ \cline{5-8} Leading                 & & Ref.
&       &    &      & Point & without \\ conf.                   & Term &
\cite{Dinh:08,Dinh:08a} &       &    &      & nucl. & Breit \\ \hline
$8s_{1/2}^1$            & (J=1/2) &        \multicolumn{6}{l}{$\rightarrow$}\\
$8p_{1/2}^1$            & (J=1/2) & 24851 & 24551 & 55 & -372 &  1299 & -51 \\
$7d_{3/2}^1$            & (J=3/2) &       & 25355 & 43 & -151 &  1714 & 135 \\
$7d_{5/2}^1$            & (J=5/2) &       & 27529 & 20 & -481 &  1661 & 152 \\
$8p_{3/2}^1$            & (J=3/2) & 38057 & 37643 & 23 & -401 &  1632 &  62 \\
$E120^{2+}$             & (J=0)   & 89931 & 89601 & 36 & -620 &  1592 &  77 \\
\hline                                     $8s_{1/2}^2$            & (J=0)   &
\multicolumn{6}{l}{$\rightarrow$}\\ $8s_{1/2}^1 8p_{1/2}^1$ & (J=0)   & 16061 &
15328 & 58 & -290 &   975 & -45 \\ $8s_{1/2}^1 8p_{1/2}^1$ & (J=1)   & 17968 &
17382 & 50 & -290 &  1015 & -30 \\ $8s_{1/2}^1 7d_{3/2}^1$ & (J=1)   & 23066 &
22337 & 30 & -145 &  1242 & 101 \\ $8s_{1/2}^1 7d_{3/2}^1$ & (J=2)   & 23231 &
22494 & 24 & -249 &  1212 & 101 \\ $8s_{1/2}^1 7d_{5/2}^1$ & (J=3)   & 23827 &
23377 & 12 & -418 &  1186 & 108 \\ $8s_{1/2}^1 8p_{3/2}^1$ & (J=2)   & 25457 &
25308 & 26 & -264 &  1186 &  42 \\ $8s_{1/2}^1 7d_{5/2}^1$ & (J=2)   & 27477 &
27652 & 15 & -401 &  1347 & 100 \\ $8s_{1/2}^1 8p_{3/2}^1$ & (J=1)   & 27685 &
28304 &  3 & -422 &  1024 &  39 \\ $8s_{1/2}^1$            & (J=1/2) & 47296 &
47633 & 11 & -458 &   906 &  40 \\ 
\hline 
\hline 
\end{tabular} 
\end{table}

We have generated the GRECP for only 10 explicitly treated electrons of E120
(i.e.\ with the $7spd,8sp,6f,5g$ GRECP components) following the
scheme \cite{Mosyagin:06amin} just to reduce unnecessary computational efforts at
the stage of correlation molecular calculations (see below);
% the
similar GRECP version was generated earlier for Ba \cite{Kozlov:97}. The results
for 10 correlated electrons (occupying the $7spd,8sp$ shells) in the
[9,11,11,9,8,7] basis set of $7{-}15s_{1/2}$, $7{-}17p_{1/2}$, $7{-}17p_{3/2}$,
$\ldots$ $6{-}12h_{11/2}$ numerical spinors localized in the same radial space
region as the $7spd,8sp$ spinors are presented in table~\ref{RECP}.  Transition
energies from the DCB/FS-RCCSD calculations with Fermi nuclear model and the
absolute errors of their reproducing in the GRECP calculations are tabulated in
the third and fourth columns. 

  Calculations with different approximations to the ``full'' GRECP operator are considered in the last four columns. If one neglects the difference between the outercore
($7sp$) and valence ($8sp$) GRECP components, two extreme
GRECP versions can be derived with the conventional semi-local RECP operator:
only valence or only outercore GRECP components acting on both the valence and
outercore electrons. These cases are referred to as the valence or core GRECP
versions, respectively \cite{Mosyagin:10a}.  One can see that the full and
valence GRECP versions are suitable for the accuracy of
2~kcal/mol\,$\approx$\,700~\cm\ whereas the core GRECP version is not. It should be
noted that the errors of neglecting the innercore correlations with the $6spd$
shells and the errors of the GRECP approximation are partly compensating each
other, thus, it additionally justifies our choice of the 10-electron GRECP for
the present molecular calculations.

The scalar-relativistic (SR) calculations, i.e.\ without spin-orbit (SO) part of
the valence GRECP operator, are presented in the seventh column.  The SO
contributions are large and should be taken into account.  The
scalar-relativistic SCF calculations followed by the FS-RCCSD calculations with
the SO part of the valence GRECP operator are presented in the last column.
These errors are comparable with the errors of the valence GRECP calculations in
the fifth column. It should be emphasized that the GRECP calculations presented in the
$4{-}6$-th columns were carried out with the SO part at both the SCF and
FS-RCCSD stages.  The computational versions used in the last two columns (the
SR~SCF calculation followed by the scalar-relativistic or fully-relativistic
coupled cluster study) are also used in the molecular correlation calculations
discussed below.

\begin{table}[!h] \caption{Transition energies (TE) from DCB/FS-RCCSD
calculations of the lowest-lying states of the E120 atom and its cations for 10
correlated electrons with Fermi nuclear model and Breit interaction are taken
into account. Absolute errors of reproducing the TE with different versions of
the GRECP calculations. All values are in cm$^{-1}$. } \label{RECP}
\begin{tabular}{llrrrrrr} 
\hline 
\hline 
   &         & TE   & \multicolumn{5}{c}{GRECP
abs.errors} \\ \cline{4-8} &         &      &      &      &      &        & val.
\\ Leading                 &         &      &      &      &      & val.   &
SR-SCF \\ conf.                   & Term    & DCB  & full & val. & core & SR &
SO-RCC \\ \hline $8s_{1/2}^1$            & (J=1/2) &
\multicolumn{6}{l}{$\rightarrow$}         \\ $8p_{1/2}^1$            & (J=1/2) &
24142 & 247 &  306 & -2761 &  9255 &  393 \\ $7d_{3/2}^1$            & (J=3/2) &
25241 & 399 &  436 & -1766 &  1316 &  358 \\ $7d_{5/2}^1$            & (J=5/2) &
27071 & 389 &  447 & -1765 &  -514 &  447 \\ $8p_{3/2}^1$            & (J=3/2) &
37194 & 326 &  392 & -1937 & -3797 &  257 \\ $E120^{2+}$             & (J=0)   &
88907 & 446 &  554 & -1701 &  -739 &  417 \\ \hline $8s_{1/2}^2$            &
(J=0)   & \multicolumn{6}{l}{$\rightarrow$}         \\ $8s_{1/2}^1 8p_{1/2}^1$ &
(J=0)   & 15012 & 266 &  310 & -2354 &  6960 & -100 \\ $8s_{1/2}^1 8p_{1/2}^1$ &
(J=1)   & 17064 & 251 &  295 & -2341 &  4908 & -171 \\ $8s_{1/2}^1 7d_{3/2}^1$ &
(J=1)   & 22207 & 366 &  401 & -1632 &   585 &  316 \\ $8s_{1/2}^1 7d_{3/2}^1$ &
(J=2)   & 22259 & 387 &  428 & -1620 &   533 &  483 \\ $8s_{1/2}^1 7d_{5/2}^1$ &
(J=3)   & 22968 & 386 &  436 & -1617 &  -176 &  484 \\ $8s_{1/2}^1 8p_{3/2}^1$ &
(J=2)   & 25009 & 253 &  297 & -1842 & -3037 &  384 \\ $8s_{1/2}^1 7d_{5/2}^1$ &
(J=2)   & 27271 & 344 &  394 & -1813 &  -262 &  291 \\ $8s_{1/2}^1 8p_{3/2}^1$ &
(J=1)   & 27834 & 307 &  351 & -1820 & -1171 &  563 \\ $8s_{1/2}^1$            &
(J=1/2) & 47120 & 333 &  385 & -1661 &  -687 &  381 \\ 
\hline
\hline 
\end{tabular}
\end{table}

To compare with the E120 transition energies, the corresponding experimental
data for Ba from Ref.~\cite{Ralchenko:11} are listed in table~\ref{Ba-exper}.
One can see that the barium excitation energies are smaller in general.  It also
indicates that E120 will possibly be more inert in general than Ba.

\begin{table}[!h] \caption{The experimental transition energies (TE) from Ref.
\cite{Ralchenko:11} for the lowest-lying states of the Ba atom and its cations.
All values are in cm$^{-1}$. } \label{Ba-exper} 
\begin{tabular}{llr} 
\hline
\hline 
 & &
Exper. \\ Leading                 &         & TE     \\ conf.  & Term    & Ref.
\cite{Ralchenko:11} \\ \hline $6s_{1/2}^1$            & (J=1/2) & $\rightarrow$
\\ $5d_{3/2}^1$            & (J=3/2) &  4874 \\ $5d_{5/2}^1$            &
(J=5/2) &  5675 \\ $6p_{1/2}^1$ & (J=1/2) & 20262 \\ $6p_{3/2}^1$            &
(J=3/2) & 21952 \\ $Ba^{2+}$ & (J=0)   & 80686 \\ \hline
$6s_{1/2}^2$ & (J=0)   & $\rightarrow$ \\ $6s_{1/2}^1 5d_{3/2}^1$ & (J=1)   &
9034 \\ $6s_{1/2}^1 5d_{3/2}^1 + 6s_{1/2}^1 5d_{5/2}^1$ & (J=2)   &  9216 \\
$6s_{1/2}^1 5d_{5/2}^1$ & (J=3)   &  9597 \\ $6s_{1/2}^1 5d_{5/2}^1 + 6s_{1/2}^1
5d_{3/2}^1$ & (J=2)   & 11395 \\ $6s_{1/2}^1 6p_{1/2}^1$ & (J=0)   & 12266 \\
$6s_{1/2}^1 6p_{1/2}^1 + 6s_{1/2}^1 6p_{3/2}^1$ & (J=1)   & 12637 \\ $6s_{1/2}^1
6p_{3/2}^1$ & (J=2)   & 13515 \\ $6s_{1/2}^1 6p_{3/2}^1 + 6s_{1/2}^1 6p_{1/2}^1$
& (J=1)   & 18060 \\ $6s_{1/2}^1$            & (J=1/2) & 42035 \\ 
\hline
\hline 
\end{tabular} \end{table}

%-----------------------------------
\subsection{Molecular calculations}

In two-component molecular relativistic calculations and high-level correlation
treatment only relatively small basis sets can be used for diatomics like
E120Au.  At the same time, rather large basis sets can be employed in
scalar-relativistic calculations of diatomic molecules.  Therefore, the
following scheme for the basis set generation was used in this work: (i) For each
atom (E120, Ba and Au), a large set of primitive Gaussian functions capable of
describing wave-functions of the ground and excited states of the corresponding
atoms was generated.  These basis sets will be referred as LBas below.
LBas(E120) and LBas(Ba) consist of $15 s-, 15 p-, 8 d-, 8 f-, 6 g-, 6 h-$type
functions, which shortly can be written as [15,15,8,8,6,6]. 
% LBas(Au) consists of $15s,\ 15p,\ ,8d,\ 8f, \ 7g$.
(ii) Then scalar relativistic CCSD calculation is performed with the large
basis set for an atom and its compound (E120, E120-H, E120-Au, etc.).
(iii) Generation of a compact basis set of contracted Gaussian functions was
performed in a manner similar to that employed for generating atomic natural
basis sets \cite{Almlof:87}: the atomic blocks from the density matrix
calculated at stage (ii) were diagonalized to yield atomic natural-like basis
set.  The functions with the largest occupation numbers were selected from these
natural basis functions. The results obtained with the given basis set
approximately reproduce those with the large basis set.  Besides, the functions
required for accurate reproducing the essentially different radial parts of the
$7p_{1/2}$ and $7p_{3/2}$ spinors, as well as the $8p_{1/2}$, $8p_{3/2}$,
$7d_{3/2}$, and $7d_{5/2}$ spinors, have also been included to the new bases.
These compact basis sets will be referred as CBas.

%scheme of molecular calculations
Finally, the following scheme to evaluate ionization potentials and dissociation
energies of molecules was employed:
(i) Calculation using two-component Fock-Space coupled cluster method with
single and double amplitudes in the CBas basis set.
(ii) Calculation of corrections on enlargement of the basis set and contribution
of triple cluster amplitudes by the scalar-relativistic coupled cluster method
with single, double and non-iterative triple cluster amplitudes, CCSD(T), using
LBas.
(iii) Calculation of corrections on higher (iterative triple and non-iterative
quadruple) cluster amplitudes by the scalar-relativistic coupled  cluster method
with single, double, triple and non-iterative quadruple cluster amplitudes,
CCSDT(Q), using CBas.  

The scheme described above was employed to calculate potential curves of the
diatomic molecules under consideration. These curves were then used to calculate
spectroscopic properties such as equilibrium internuclear distances, vibrational
constants, etc.

%-------------------------------- --------------------------------
\section{Results and discussions}

%---------------------------------
\subsection{Ionization potentials}

Some properties of E120, of which the first (IP1) and second (IP2) 
ionization potentials are examples, are considered here in comparison 
with the corresponding properties of Ba.
As was described above, the two-component FS-RCCSD method 
was used to calculate the main contributions to IP1 and IP2. 
These calculations were 
performed in compact basis sets, CBas, consisting of 
$5 s, 6 p, 4 d,$ and $3 f$ 
functions for E120 and of $5 s, 5 p, 3 d,$ and $3 f$ for Ba.
% ат.данные IP1=47505 IP2=89461
The computed values for E120 are IP1$=47236$ \cm\ and IP2$=89061$ \cm, 
while for Ba IP1$=42340$ \cm\ and IP2$=80326$ \cm. Contributions 
from enlargement of the basis set up to $15 s, 15 p, 8 d, 8 f$ and $6 h$
([15,15,8,8,6,6]) functions and non-iterative triple cluster amplitudes
for E120 are -462 \cm\ for IP1 and 31 \cm\ for IP2 
  \newcounter{fnnumber}
  \footnote{We have also included here the  corrections arising when    
   going from the Fock-Space coupled clusters to the single-reference 
   coupled clusters.}%
   \setcounter{fnnumber}{\thefootnote}%
. For Ba these values are -579 \cm and -65 \cm. Contributions of higher cluster
amplitudes estimated using CCSDT(Q) method and [15,15,8,8,6] basis set ($h$
functions were excluded) are negligible (less than 30 \cm).  The final values of
the ionization potentials for E120 and Ba are given in table \ref{TFinal}
together with the corresponding experimental values for Ba. 

The theoretical uncertainty of the ionization potentials of E120 can be estimated from the corresponding atomic calculations 
% Надо отсчитывать от 28-эл. DCB и затем брать SR-SCF SO-RCCSD
% IP1=-620+417=-203 IP2=-458+381=-77
(tables \ref{All-el} and \ref{RECP}) and is suggested to be within
1~kcal/mol$\approx$350~\cm.

%Ca:   49 305.95   95 751.87 
%Sr:   45 932.204  88 965.18 
%Ba:   42 034.910  80 686.25 
%Ra:   42 573.36   81 842.31 
%E120: 47236      89061

%-------------------------------------------------
\subsection{X-H dissociation energies, X=E120, Ba}

In order to estimate the stability of compounds of E120 compared to those of Ba
we have first considered the dissociation energies of the corresponding hydrides
% for which high-level correlation calculations can be performed. 
since the experimental data are available for BaH \cite{Huber:79}).  Another
often considered characteristic of SHE is the dissociation energy of its
fluorides, X-F. However, the X-F bonding is not so illustrative qualitatively
because almost all the elements (except light noble gases) are known to react in
a fluorine atmosphere yielding rather stable fluorides.  At the same time the
dimer systems such as Ba$_2$, Hg$_2$, Xe$_2$, E112$_2$ are all the van der Waals
systems with small dissociation energies.  By contrast, the ground state of the
XeH molecule is not observed in the gas phase, whereas BaH was obtained and
characterized \cite{Knight:71, Walker:93}.

To calculate E120H and BaH, the scheme similar to that for the calculation of
the ionization potentials was used.  Compact basis sets for E120, Ba and H were
[5,6,4,2], [5,5,3,2] and [4,3,1], respectively. Large basis sets for E120 and Ba
were [15,15,8,8,6], i.e.\ without $h$-functions.  The aug-cc-pvqz
\cite{Dunning:89} basis set was used as the large basis set for H.  To exclude
the basis set superposition errors, the diatomic molecules and atoms were
calculated in the same two-center basis, i.e.\ the counterpoise corrections
\cite{Boys:70} were used.

  The dissociation energy of E120H calculated within FS-RCCSD using the
  CBas(E120) and CBas(H) basis sets is 8117~\cm, while the correction on the
  large basis and triple cluster amplitudes is -64 \cm.  For BaH the former
  contribution is 16846 \cm, while the correction is -215 \cm.

The final values for the dissociation energies and other calculated
spectroscopic properties of E120H and BaH are given in table \ref{TFinal}.  The equilibrium internuclear distance in BaH is on 0.14~\AA\ shorter than that in 
E120H\footnote{Note that similar relation between internuclear distances takes place in E120H$^+$ and BaH$^+$ that is considered and discussed in \cite{Thierfelder:09}.}\addtocounter{footnote}{-1}.

The theoretical uncertainty of the dissociation energy of E120H is estimated to
be 500~\cm.

%--------------
It follows from table~\ref{TFinal} that the E120-H bond is significantly weaker
than that in BaH. Partly it can be explained by the presence of lower-lying
excited states in the case of Ba and its cation (see table~\ref{Ba-exper}).
Contribution of these states to the BaH chemical bond leads to its additional
stabilization. In turn, the excited states in the case of E120 are lying
significantly higher (see table~\ref{RECP});
%   another order of the excited levels in the case of E120 is mainly explained
%   by the relativistic stabilization of %its $s$ and $p_{1/2}$ states.

To check this viewpoint two series of the FS-RCCSD calculations of the BaH and
E120H dissociation energies were performed:
(i) with $d-$ and $f-$type basis functions included in the basis set (CBas);
(ii) without $d$ and $f$ functions.  The latter calculation prevents
participating the $d$~orbitals of Ba and E120 in chemical bonding of their
hydrides.  Note, however, that it is just test calculation because exclusion of
$d$~basis functions also prevents, e.g., correlation of $5p-$electrons of Ba and
$7p-$electrons of E120 into these states. The results are given in
table~\ref{THydridesBonding}.

\begin{table}[!h]
\caption{Calculated dissociation energy of BaH and E120H using
         scalar-relativistic Hartree-Fock (HF) or FS-RCCSD levels of theory in
         CBas basis set and CBas with excluded $d-,f-$ functions on Ba and E120
         (CBasNoD) }

\label{THydridesBonding}
\begin{tabular}{lllll}
\hline 
\hline 
Basis & CBas & • & CBasNoD& • \\ 
\hline 
Method & HF & FS-RCCSD & HF & FS-RCCSD \\ 
\hline 
$D_e$(BaH),~ \cm & 13605 & 16846 & 8634 & 11699 \\ 
%\hline 
$D_e$(E120H),~ \cm & 2642 & 8117 & 61 & 6366 \\ 
\hline 
\hline 
\end{tabular} 
\end{table}

It follows from table~\ref{THydridesBonding} that $d$ functions (and higher
harmonics) significantly contribute to bonding of the monohydrides under
consideration. At the FS-RCCSD level of theory, this contribution is 5147~\cm\
to the Ba-H bond, while is only 1751~\cm\ for E120H. These values confirm the
qualitative discussion above based on the atomic transitions.  However, even
without $d$ basis functions the BaH bond energy at the FS-RCCSD level is twice
stronger than that in E120H.  This observation is in a qualitative agreement
with the fact that the states with the valence $sp$ ($p$) configurations in E120
(E120$^+$) lies higher than the corresponding states in Ba ($Ba^+$). It should
be noted that the bonding in E120H is in essence due to the correlation effects
when the CBasNoD basis set is used.

%--------------------------------------------------
\subsection{X-Au dissociation energies, X=E120, Ba}

As the first stage of modeling interaction of E120 with gold surface, the
simplest comparative model, E120Au vs.\ BaAu, is considered here.

The 19-electron GRECP was used for Au. Thus, 29 electrons were treated in the
correlation calculation. For E120Au, the relativistic two-component FS-RCCSD
calculation in the Au[7,7,4,2] and E120[5,5,2,1] basis sets gives the
dissociation energy of 11535 \cm. Correction on the larger basis set
([15,15,8,8,7] for Au and [15,15,8,4] for E120) is 837 \cm, contribution of non-iterative triple cluster amplitudes 
   \footnotemark[\thefootnote]
is 2262 \cm\ and correction on higher amplitudes (calculated as the difference
between CCSDT(Q) and CCSD(T) energies in the compact Au[6,6,4,2,1] and
E120[6,5,2,1] basis sets) is less than 100 \cm.  Similar calculations were
performed for BaAu. The final calculated values are given in table \ref{TFinal}.
The equilibrium distance for BaAu is on 0.12 \AA\ shorter than that in E120Au.

The theoretical uncertainty of the dissociation energy of E120Au is estimated to be 1000~\cm.

\begin{table}[!h]
\caption{
Calculated properties of E120 in comparison with Ba:
the first and second ionization potentials
(IP1 and IP2), dissociation energies ($D_e$), equilibrium internuclear distances ($R_e$), harmonic frequency ($w_e$), vibrational anharmonicity ($w_e x_e$) of hydrides and aurides (X-Au).
}
\label{TFinal}
\begin{tabular}{llll}
\hline 
\hline 
 & E120-calc & Ba-calc  & Ba-exp \\ 
\hline 
IP1, \cm & 47046 & 41932 & 42035 \cite{Moore:58} \\ 

IP2, \cm & 89286 & 80442 & 80686 \cite{Moore:58} \\ 
\hline 
$D_e$(X-H), \cm & 8053 & 16631 & 
$\leq$
%end ls
16308\cite{Huber:79} \\ 
$R_e$(X-H), \AA & 2.38 & 2.24 & 2.23  \cite{Knight:71} \\ 
$w_e$(X-H), \cm      &   1070 & 1158 & 1168  \cite{Knight:71} \\ 
%\hline 
$w_e x_e$(X-H), \cm &    20.1 & 14.1 & 14.5  \cite{Knight:71} \\
\hline 
$D_e$(X-Au), \cm & 14537 & 22909 & - \\ 

$R_e$(X-Au), \AA & 3.03 & 2.91 & - \\ 

$w_e$(X-Au), \cm & 100    &  125   & 129  \cite{Huber:79} \\ 
%\hline 
$w_e x_e$(X-Au), \cm &  0.13   & 0.16 & 0.18  \cite{Huber:79} \\
\hline 
\hline 
\end{tabular} 
\end{table}

%----------------------
\subsection{Conclusion}

Properties of E120 and its compounds are considered in comparison with their Ba
analogues. The monohydride and monoauride of E120 are found to be less stable
than the corresponding analogues of the Ba compounds.
%  Such a trend in decreasing the dissociation energies with E120 compared to
%  lighter homologue is mainly due to increasing the excitation (activation)
%  energies in E120 compared to Ba that, in turn, is caused by the relativistic
%  stabilization of the valence s and p1/2 shells in superheavy elements.
%  However, such a  decreasing in the bonding energies with E120 is not so
%  drammatic as is in the cases of Cn and E114
 Nevertheless, E120 can be rather considered as a ``typical'' representative of
 the second group.
% having no featires of a noble gas element.

\subsection{ACKNOWLEDGMENTS}
We are grateful to Prof.\ A.V.\ Zaitsevskii for valuable discussions.  This work
is supported by the RFBR grant 11-03-12155-ofi-m-2011.  L.S.\ is grateful to the
Dmitry Zimin ``Dynasty'' Foundation.  The molecular calculations were performed at the Supercomputer ``Lomonosov''.

%\bibliographystyle{./bib/apsrev}
%\bibliography{bib/JournAbbr,bib/SkripnikovLib,bib/Titov,bib/TitovLib,bib/Kaldor}

\begin{thebibliography}{25}
\expandafter\ifx\csname natexlab\endcsname\relax\def\natexlab#1{#1}\fi
\expandafter\ifx\csname bibnamefont\endcsname\relax
  \def\bibnamefont#1{#1}\fi
\expandafter\ifx\csname bibfnamefont\endcsname\relax
  \def\bibfnamefont#1{#1}\fi
\expandafter\ifx\csname citenamefont\endcsname\relax
  \def\citenamefont#1{#1}\fi
\expandafter\ifx\csname url\endcsname\relax
  \def\url#1{\texttt{#1}}\fi
\expandafter\ifx\csname urlprefix\endcsname\relax\def\urlprefix{URL }\fi
\providecommand{\bibinfo}[2]{#2}
\providecommand{\eprint}[2][]{\url{#2}}

\bibitem[{\citenamefont{Oganessian}(2009)}]{Oganessian:09a}
\bibinfo{author}{\bibfnamefont{Y.}~\bibnamefont{Oganessian}},
  \bibinfo{journal}{Eur.\ Phys.\ J.\ A} \textbf{\bibinfo{volume}{42}},
  \bibinfo{pages}{361} (\bibinfo{year}{2009}).

\bibitem[{\citenamefont{Oganessian et~al.}(2010)\citenamefont{Oganessian,
  Abdullin, Bailey, Benker, Bennett, Dmitriev, Ezold, Hamilton, Henderson,
  Itkis et~al.}}]{Oganessian:10a}
\bibinfo{author}{\bibfnamefont{Y.~T.} \bibnamefont{Oganessian}},
  \bibinfo{author}{\bibfnamefont{F.~S.} \bibnamefont{Abdullin}},
  \bibinfo{author}{\bibfnamefont{P.~D.} \bibnamefont{Bailey}},
  \bibinfo{author}{\bibfnamefont{D.~E.} \bibnamefont{Benker}},
  \bibinfo{author}{\bibfnamefont{M.~E.} \bibnamefont{Bennett}},
  \bibinfo{author}{\bibfnamefont{S.~N.} \bibnamefont{Dmitriev}},
  \bibinfo{author}{\bibfnamefont{J.~G.} \bibnamefont{Ezold}},
  \bibinfo{author}{\bibfnamefont{J.~H.} \bibnamefont{Hamilton}},
  \bibinfo{author}{\bibfnamefont{R.~A.} \bibnamefont{Henderson}},
  \bibinfo{author}{\bibfnamefont{M.~G.} \bibnamefont{Itkis}},
  \bibnamefont{et~al.}, \bibinfo{journal}{Phys.\ Rev.\ Lett.}
  \textbf{\bibinfo{volume}{104}}, \bibinfo{pages}{142502}
  (\bibinfo{year}{2010}).

\bibitem[{\citenamefont{Stavsetra et~al.}(2009)\citenamefont{Stavsetra,
  Gregorich, Dvorak, Ellison, Dragojevi\ifmmode~\acute{c}\else \'{c}\fi{},
  Garcia, and Nitsche}}]{Stavsetra:09}
\bibinfo{author}{\bibfnamefont{L.}~\bibnamefont{Stavsetra}},
  \bibinfo{author}{\bibfnamefont{K.~E.} \bibnamefont{Gregorich}},
  \bibinfo{author}{\bibfnamefont{J.}~\bibnamefont{Dvorak}},
  \bibinfo{author}{\bibfnamefont{P.~A.} \bibnamefont{Ellison}},
  \bibinfo{author}{\bibfnamefont{I.}~\bibnamefont{Dragojevi\ifmmode~\acute{c}\%
else \'{c}\fi{}}}, \bibinfo{author}{\bibfnamefont{M.~A.} \bibnamefont{Garcia}},
  \bibnamefont{and} \bibinfo{author}{\bibfnamefont{H.}~\bibnamefont{Nitsche}},
  \bibinfo{journal}{Phys. Rev. Lett.} \textbf{\bibinfo{volume}{103}},
  \bibinfo{pages}{132502} (\bibinfo{year}{2009}).

\bibitem[{\citenamefont{Gates et~al.}(2011)\citenamefont{Gates, {D\"ullmann},
  {Sch\"adel}, Yakushev, T\"urler, Eberhardt, Kratz, Ackermann, Andersson,
  Block et~al.}}]{Gates:11a}
\bibinfo{author}{\bibfnamefont{J.~M.} \bibnamefont{Gates}},
  \bibinfo{author}{\bibfnamefont{C.~E.} \bibnamefont{{D\"ullmann}}},
  \bibinfo{author}{\bibfnamefont{M.}~\bibnamefont{{Sch\"adel}}},
  \bibinfo{author}{\bibfnamefont{A.}~\bibnamefont{Yakushev}},
  \bibinfo{author}{\bibfnamefont{A.}~\bibnamefont{T\"urler}},
  \bibinfo{author}{\bibfnamefont{K.}~\bibnamefont{Eberhardt}},
  \bibinfo{author}{\bibfnamefont{J.~V.} \bibnamefont{Kratz}},
  \bibinfo{author}{\bibfnamefont{D.}~\bibnamefont{Ackermann}},
  \bibinfo{author}{\bibfnamefont{L.-L.} \bibnamefont{Andersson}},
  \bibinfo{author}{\bibfnamefont{M.}~\bibnamefont{Block}},
  \bibnamefont{et~al.}, \bibinfo{journal}{Phys.\ Rev.\ C}
  \textbf{\bibinfo{volume}{83}}, \bibinfo{pages}{054618}
  (\bibinfo{year}{2011}).

\bibitem[{\citenamefont{Dinh et~al.}(2008{\natexlab{a}})\citenamefont{Dinh,
  Dzuba, Flambaum, and Ginges}}]{Dinh:08}
\bibinfo{author}{\bibfnamefont{T.~H.} \bibnamefont{Dinh}},
  \bibinfo{author}{\bibfnamefont{V.~A.} \bibnamefont{Dzuba}},
  \bibinfo{author}{\bibfnamefont{V.~V.} \bibnamefont{Flambaum}},
  \bibnamefont{and} \bibinfo{author}{\bibfnamefont{J.~S.~M.}
  \bibnamefont{Ginges}}, \bibinfo{journal}{Phys. Rev. A}
  \textbf{\bibinfo{volume}{78}}, \bibinfo{pages}{054501}
  (\bibinfo{year}{2008}{\natexlab{a}}).

\bibitem[{\citenamefont{Schwerdtfeger and Seth}(1998)}]{Schwerdtfeger:98aa}
\bibinfo{author}{\bibfnamefont{P.}~\bibnamefont{Schwerdtfeger}}
  \bibnamefont{and} \bibinfo{author}{\bibfnamefont{M.}~\bibnamefont{Seth}}, in
  \emph{\bibinfo{booktitle}{Encyclopedia of Computational Chemistry}}, edited
  by \bibinfo{editor}{\bibfnamefont{P.}~\bibnamefont{{von Ragu\'e Schleyer et
  al.}}} (\bibinfo{organization}{Wiley}, \bibinfo{address}{New York},
  \bibinfo{year}{1998}), vol.~\bibinfo{volume}{4}, pp.
  \bibinfo{pages}{2480--2499}.

\bibitem[{\citenamefont{Kaldor et~al.}(2004)\citenamefont{Kaldor, Eliav, and
  Landau}}]{Kaldor:04ba}
\bibinfo{author}{\bibfnamefont{U.}~\bibnamefont{Kaldor}},
  \bibinfo{author}{\bibfnamefont{E.}~\bibnamefont{Eliav}}, \bibnamefont{and}
  \bibinfo{author}{\bibfnamefont{A.}~\bibnamefont{Landau}}, in
  \emph{\bibinfo{booktitle}{Recent Advances in Relativistic Molecular Theory}},
  edited by \bibinfo{editor}{\bibfnamefont{K.}~\bibnamefont{Hirao}}
  \bibnamefont{and} \bibinfo{editor}{\bibfnamefont{Y.}~\bibnamefont{Ishikawa}}
  (\bibinfo{organization}{World Scientific}, \bibinfo{address}{Singapore},
  \bibinfo{year}{2004}), p. \bibinfo{pages}{283}.

\bibitem[{\citenamefont{Eichler et~al.}(2007)\citenamefont{Eichler, Aksenov,
  Belozerov et~al.}}]{Eichler:07a}
\bibinfo{author}{\bibfnamefont{R.}~\bibnamefont{Eichler}},
  \bibinfo{author}{\bibfnamefont{N.~V.} \bibnamefont{Aksenov}},
  \bibinfo{author}{\bibfnamefont{A.~V.} \bibnamefont{Belozerov}},
  \bibnamefont{et~al.}, \bibinfo{journal}{Nature}
  \textbf{\bibinfo{volume}{447}}, \bibinfo{pages}{72} (\bibinfo{year}{2007}).

\bibitem[{\citenamefont{Mosyagin et~al.}(2010)\citenamefont{Mosyagin,
  Zaitsevskii, and Titov}}]{Mosyagin:10a}
\bibinfo{author}{\bibfnamefont{N.~S.} \bibnamefont{Mosyagin}},
  \bibinfo{author}{\bibfnamefont{A.~V.} \bibnamefont{Zaitsevskii}},
  \bibnamefont{and} \bibinfo{author}{\bibfnamefont{A.~V.} \bibnamefont{Titov}},
  \bibinfo{journal}{Review of Atomic and Molecular Physics}
  \textbf{\bibinfo{volume}{1}}, \bibinfo{pages}{63} (\bibinfo{year}{2010}).

\bibitem[{\citenamefont{Kaldor et~al.}()\citenamefont{Kaldor, Eliav, and
  Landau}}]{MolRCCSD}
\bibinfo{author}{\bibfnamefont{U.}~\bibnamefont{Kaldor}},
  \bibinfo{author}{\bibfnamefont{E.}~\bibnamefont{Eliav}}, \bibnamefont{and}
  \bibinfo{author}{\bibfnamefont{A.}~\bibnamefont{Landau}},
  \bibinfo{note}{program package for calculation of molecules by the
  relativistic {F}ock-space coupled-cluster method}.

\bibitem[{\citenamefont{Stanton et~al.}(2011)\citenamefont{Stanton, Gauss,
  Harding, Szalay et~al.}}]{CFOUR}
\bibinfo{author}{\bibfnamefont{J.~F.} \bibnamefont{Stanton}},
  \bibinfo{author}{\bibfnamefont{J.}~\bibnamefont{Gauss}},
  \bibinfo{author}{\bibfnamefont{M.~E.} \bibnamefont{Harding}},
  \bibinfo{author}{\bibfnamefont{P.~G.} \bibnamefont{Szalay}},
  \bibnamefont{et~al.} (\bibinfo{year}{2011}), \bibinfo{note}{{\sc cfour}: a
  program package for performing high-level quantum chemical calculations on
  atoms and molecules, {http://www.cfour.de} .}

\bibitem[{\citenamefont{K\'{a}llay and Surj\'{a}n}(2001)}]{Kallay:1}
\bibinfo{author}{\bibfnamefont{M.}~\bibnamefont{K\'{a}llay}} \bibnamefont{and}
  \bibinfo{author}{\bibfnamefont{P.~R.} \bibnamefont{Surj\'{a}n}},
  \bibinfo{journal}{J.\ Chem.\ Phys.} \textbf{\bibinfo{volume}{115}},
  \bibinfo{pages}{2945} (\bibinfo{year}{2001}).

\bibitem[{\citenamefont{K\'{a}llay et~al.}(2003)\citenamefont{K\'{a}llay,
  Gauss, and Szalay}}]{Kallay:3}
\bibinfo{author}{\bibfnamefont{M.}~\bibnamefont{K\'{a}llay}},
  \bibinfo{author}{\bibfnamefont{J.}~\bibnamefont{Gauss}}, \bibnamefont{and}
  \bibinfo{author}{\bibfnamefont{P.~G.} \bibnamefont{Szalay}},
  \bibinfo{journal}{J.\ Chem.\ Phys.} \textbf{\bibinfo{volume}{119}},
  \bibinfo{pages}{2991} (\bibinfo{year}{2003}).

\bibitem[{\citenamefont{Dinh et~al.}(2008{\natexlab{b}})\citenamefont{Dinh,
  Dzuba, Flambaum, and Ginges}}]{Dinh:08a}
\bibinfo{author}{\bibfnamefont{T.~H.} \bibnamefont{Dinh}},
  \bibinfo{author}{\bibfnamefont{V.~A.} \bibnamefont{Dzuba}},
  \bibinfo{author}{\bibfnamefont{V.~V.} \bibnamefont{Flambaum}},
  \bibnamefont{and} \bibinfo{author}{\bibfnamefont{J.~S.~M.}
  \bibnamefont{Ginges}}, \bibinfo{journal}{Phys. Rev. A}
  \textbf{\bibinfo{volume}{78}}, \bibinfo{pages}{022507}
  (\bibinfo{year}{2008}{\natexlab{b}}).

\bibitem[{\citenamefont{Mosyagin et~al.}(2006)\citenamefont{Mosyagin, Petrov,
  Titov, and Tupitsyn}}]{Mosyagin:06amin}
\bibinfo{author}{\bibfnamefont{N.~S.} \bibnamefont{Mosyagin}},
  \bibinfo{author}{\bibfnamefont{A.~N.} \bibnamefont{Petrov}},
  \bibinfo{author}{\bibfnamefont{A.~V.} \bibnamefont{Titov}}, \bibnamefont{and}
  \bibinfo{author}{\bibfnamefont{I.~I.} \bibnamefont{Tupitsyn}},
  \bibinfo{journal}{Progr.\ Theor.\ Chem.\ Phys.}
  \textbf{\bibinfo{volume}{B~15}}, \bibinfo{pages}{229} (\bibinfo{year}{2006}).

\bibitem[{\citenamefont{Kozlov et~al.}(1997)\citenamefont{Kozlov, Titov,
  Mosyagin, and Souchko}}]{Kozlov:97}
\bibinfo{author}{\bibfnamefont{M.~G.} \bibnamefont{Kozlov}},
  \bibinfo{author}{\bibfnamefont{A.~V.} \bibnamefont{Titov}},
  \bibinfo{author}{\bibfnamefont{N.~S.} \bibnamefont{Mosyagin}},
  \bibnamefont{and} \bibinfo{author}{\bibfnamefont{P.~V.}
  \bibnamefont{Souchko}}, \bibinfo{journal}{Phys.\ Rev.\ A}
  \textbf{\bibinfo{volume}{56}}, \bibinfo{pages}{R3326} (\bibinfo{year}{1997}).

\bibitem[{\citenamefont{Ralchenko et~al.}(2011)\citenamefont{Ralchenko,
  Kramida, and Reader}}]{Ralchenko:11}
\bibinfo{author}{\bibfnamefont{Y.}~\bibnamefont{Ralchenko}},
  \bibinfo{author}{\bibfnamefont{A.}~\bibnamefont{Kramida}}, \bibnamefont{and}
  \bibinfo{author}{\bibfnamefont{J.}~\bibnamefont{Reader}},
  \emph{\bibinfo{title}{{NIST} Atomic Spectra Database (ver. 4.1.0)}}
  (\bibinfo{publisher}{National Institute of Standards and Technology},
  \bibinfo{address}{Gaithersburg, MD}, \bibinfo{year}{2011}),
  \bibinfo{note}{http://physics.nist.gov/asd}.

\bibitem[{\citenamefont{{Alml\"of} and Taylor}(1987)}]{Almlof:87}
\bibinfo{author}{\bibfnamefont{J.}~\bibnamefont{{Alml\"of}}} \bibnamefont{and}
  \bibinfo{author}{\bibfnamefont{P.~R.} \bibnamefont{Taylor}},
  \bibinfo{journal}{J.\ Chem.\ Phys.} \textbf{\bibinfo{volume}{86}},
  \bibinfo{pages}{4070} (\bibinfo{year}{1987}).

\bibitem[{\citenamefont{Huber and Herzberg}(1979)}]{Huber:79}
\bibinfo{author}{\bibfnamefont{K.~P.} \bibnamefont{Huber}} \bibnamefont{and}
  \bibinfo{author}{\bibfnamefont{G.}~\bibnamefont{Herzberg}},
  \emph{\bibinfo{title}{Constants of Diatomic Molecules}}
  (\bibinfo{publisher}{Van Nostrand-Reinhold}, \bibinfo{address}{New York},
  \bibinfo{year}{1979}).

\bibitem[{\citenamefont{Knight et~al.}(1971)\citenamefont{Knight, Easley, and
  Weltner}}]{Knight:71}
\bibinfo{author}{\bibfnamefont{L.~B.} \bibnamefont{Knight}},
  \bibinfo{author}{\bibfnamefont{W.~C.} \bibnamefont{Easley}},
  \bibnamefont{and} \bibinfo{author}{\bibfnamefont{W.}~\bibnamefont{Weltner}},
  \bibinfo{journal}{J.\ Chem.\ Phys.} \textbf{\bibinfo{volume}{54}},
  \bibinfo{pages}{322} (\bibinfo{year}{1971}).

\bibitem[{\citenamefont{Walker et~al.}(1993)\citenamefont{Walker, Hedderich,
  and Bernath}}]{Walker:93}
\bibinfo{author}{\bibfnamefont{K.~A.} \bibnamefont{Walker}},
  \bibinfo{author}{\bibfnamefont{H.~G.} \bibnamefont{Hedderich}},
  \bibnamefont{and} \bibinfo{author}{\bibfnamefont{P.~F.}
  \bibnamefont{Bernath}}, \bibinfo{journal}{Mol.\ Phys.}
  \textbf{\bibinfo{volume}{78}}, \bibinfo{pages}{577} (\bibinfo{year}{1993}).

\bibitem[{\citenamefont{{Dunning, Jr}}(1989)}]{Dunning:89}
\bibinfo{author}{\bibfnamefont{T.~H.} \bibnamefont{{Dunning, Jr}}},
  \bibinfo{journal}{J.\ Chem.\ Phys.} \textbf{\bibinfo{volume}{90}},
  \bibinfo{pages}{1007} (\bibinfo{year}{1989}).

\bibitem[{\citenamefont{Boys and Bernardi}(1970)}]{Boys:70}
\bibinfo{author}{\bibfnamefont{S.~F.} \bibnamefont{Boys}} \bibnamefont{and}
  \bibinfo{author}{\bibfnamefont{F.}~\bibnamefont{Bernardi}},
  \bibinfo{journal}{Molecular Physics} \textbf{\bibinfo{volume}{19}},
  \bibinfo{pages}{553} (\bibinfo{year}{1970}).

\bibitem[{\citenamefont{Moore}(1958)}]{Moore:58}
\bibinfo{author}{\bibfnamefont{C.~E.} \bibnamefont{Moore}},
  \emph{\bibinfo{title}{Atomic Energy Levels}}, vol. \bibinfo{volume}{1-3}
  (\bibinfo{publisher}{Natl. Bur. Stand. (US), Circ. No. 467},
  \bibinfo{address}{Washington}, \bibinfo{year}{1958}),
  \bibinfo{note}{url=http://www.nist.gov/srd/nsrds.cfm}.

\bibitem[{\citenamefont{Thierfelder et~al.}(2009)\citenamefont{Thierfelder,
  Schwerdtfeger, Koers, Borschevsky, and Fricke}}]{Thierfelder:09}
\bibinfo{author}{\bibfnamefont{C.}~\bibnamefont{Thierfelder}},
  \bibinfo{author}{\bibfnamefont{P.}~\bibnamefont{Schwerdtfeger}},
  \bibinfo{author}{\bibfnamefont{A.}~\bibnamefont{Koers}},
  \bibinfo{author}{\bibfnamefont{A.}~\bibnamefont{Borschevsky}},
  \bibnamefont{and} \bibinfo{author}{\bibfnamefont{B.}~\bibnamefont{Fricke}},
  \bibinfo{journal}{Phys.\ Rev.\ A} \textbf{\bibinfo{volume}{80}},
  \bibinfo{pages}{022501} (\bibinfo{year}{2009}).

\end{thebibliography}

\end{document}